\documentclass{PoS}
\usepackage{graphicx}
\PoS{PoS(LAT2005)220}

\title{B semileptonic decays with 2+1 dynamical quark flavors }

\ShortTitle{}

\author{\speaker{Emel Gulez}$^a$, Christine Davies$^b$, Alan Gray$^a$, Peter Lepage$^c$, Junko Shigemitsu$^a$, Matthew Wingate$^d$\\
$^a$ Physics Department, The Ohio State University, Columbus, OH 43210, USA.\\
$^b$ Department of Physics and Astronomy, University of Glasgow, Glasgow, G12 8QQ, UK\\
$^c$ LEPP, Cornell University, Ithaca, NY 14853, USA. \\
$^d$ Institute for Nuclear Theory, University of Washington, Seattle, WA 98195, USA.}

\abstract{We study semileptonic B decays, using MILC dynamical configurations with $N_f=2+1$. NRQCD heavy and AsqTad light quark actions are used. We obtain the semileptonic form factors $f_+(q^2)$ and $f_0(q^2)$ in the chiral limit.}

\FullConference{XXIIIrd International Symposium on Lattice Field Theory\\
25-30 July 2005\\
Trinity College, Dublin, Ireland}

\begin{document}

\section{Introduction}

This work is a continuation of the study of semileptonic B decays by the HPQCD collaboration \cite{lastyear}, which was announced last year. Since then, we have included all $1/M$ corrections for the temporal and spatial parts of the heavy-light electroweak currents $|V_0|$ and $|V_k|$. Previously, only the corrections that are lowest order in $1/M$ were included. Also, last year our data was on one ensemble, now we have more ensembles, listed in \cite{advisortalk} and hence more full QCD data points in addition to partially quenched results. Moreover, we have started implementing and examining the effects of staggered chiral perturbation theory fits using formulas by Aubin and Bernard \cite{xpt}.
\section{Current Matching}

We constructed the heavy-light currents in terms of lattice operators so that they accurately represent the continuum heavy-light currents up to the desired order. We picked a process where the operator to be matched is relevant. The corresponding scattering amplitude was matched between lattice and continuum through one loop in perturbation theory. The result is independent of the process picked to do the matching. Our process is a heavy quark scattering into a light quark through the electroweak current. The corresponding Feynman diagram is calculated through one loop in continuum perturbation theory; then the result is expanded in powers of $1/M$. The terms in the expansion are identified as the continuum analogs of the necessary lattice operators. These lattice currents are:

Temporal Currents:
\begin{eqnarray}
 J^{(0)}_{0}(x) & = & \bar q(x) \,\Gamma_0\, Q(x),  \nonumber\\
\label{j1}
 J^{(1)}_{0}(x) & = & \frac{-1}{2 \,(aM_0)} \bar q(x)
    \,\Gamma_0\,\mbox{\boldmath$\gamma\!\cdot\!\nabla$} \, Q(x), \nonumber\\
\label{j2}
 J^{(2)}_{0}(x) & = & \frac{-1}{2 \,(aM_0)}  \bar q(x)
    \,\mbox{\boldmath$\gamma\!\cdot\!\overleftarrow{\nabla}$}
    \,\gamma_0\ \Gamma_0\, Q(x).\nonumber
\end{eqnarray}

Spatial Currents:
\begin{eqnarray}
 J^{(0)}_{k}(x) & = & \bar q(x) \,\Gamma_k\, Q(x),  \nonumber\\
 J^{(1)}_{k}(x) & = & \frac{-1}{2 \,(aM_0)} \bar q(x)
    \,\Gamma_k\,\mbox{\boldmath$\gamma\!\cdot\!\nabla$} \, Q(x), \nonumber\\
 J^{(2)}_{k}(x) & = & \frac{-1}{2 \,(aM_0)}  \bar q(x)
    \,\mbox{\boldmath$\gamma\!\cdot\!\overleftarrow{\nabla}$}
    \,\gamma_0\ \Gamma_k\, Q(x),\nonumber\\
 J^{(3)}_{k}(x) & = & \frac{-1}{2 \,(aM_0)} \bar q(x)
    \,\Gamma_t\,\mbox{\boldmath$\nabla_k$} \, Q(x), \nonumber\\
 J^{(4)}_{k}(x) & = & \frac{1}{2 \,(aM_0)}  \bar q(x)
    \,\mbox{\boldmath$\overleftarrow{\nabla}_k$}
    \,\gamma_t\ \Gamma_k\, Q(x)\nonumber,
\end{eqnarray}
where $\Gamma_\mu = \gamma_\mu$ or $\gamma_5  \gamma_\mu$, and
$\Gamma_t=1$ or  $\gamma_5$.

Because of the symmetries of our action, axial vector and vector currents give the same results. We used this as a further check in our matching calculations.

The continuum currents are a linear combination of the lattice currents through $O(\alpha_s/M)$:

\begin{eqnarray}
\langle V_0\rangle_{cont}&=&(1+\alpha_S\rho_0)\langle J_0^{(0)}\rangle+(1+\alpha_S\rho_1)\langle J_0^{(1),sub}\rangle+\alpha_S\rho_2 \langle J_0^{(2),sub}\rangle,\nonumber\\
\langle V_k\rangle_{cont}&=&(1+\alpha_S\rho_0)\langle J_k^{(0)}\rangle+(1+\alpha_S\rho_1)\langle J_k^{(1),sub}\rangle + \alpha_S\rho_2 \langle J_k^{(2),sub}\rangle\nonumber
  +\alpha_S\rho_3 \langle J_k^{(3)}\rangle+\alpha_S\rho_4 \langle J_k^{(4)}\rangle,\nonumber
\end{eqnarray}
where $J_0^{(i),sub}=J_0^{(i)}-\alpha_S\xi_{i0}J_0^{(0)}$. The subtracted currents $J_\mu^{(i),sub}$ are more physical, as they have power law contributions subtracted out through $O(\alpha_S/aM)$.

Last year, our calculation was correct through $O(\alpha_s)$: it only included the first term, $(1+\alpha_S\rho_0)\langle J_0^{(0)}\rangle$. We now take into account all the $1/M$ terms in the above expression. To this end, we calculated both the nonperturbative matrix elements $\langle J \rangle$ and the perturbative coefficients $\rho_i$. The perturbative calculations are described in \cite{me}.
 
The difference between matching at $O(\alpha_s/M)$ and at $O(\alpha_s)$ is found to be very small. This difference is plotted for one of the currents, $J_0^{(1)}$ in Fig.1.  
\vspace{.6in}
\begin{figure}[h]
\includegraphics[width=14.cm]{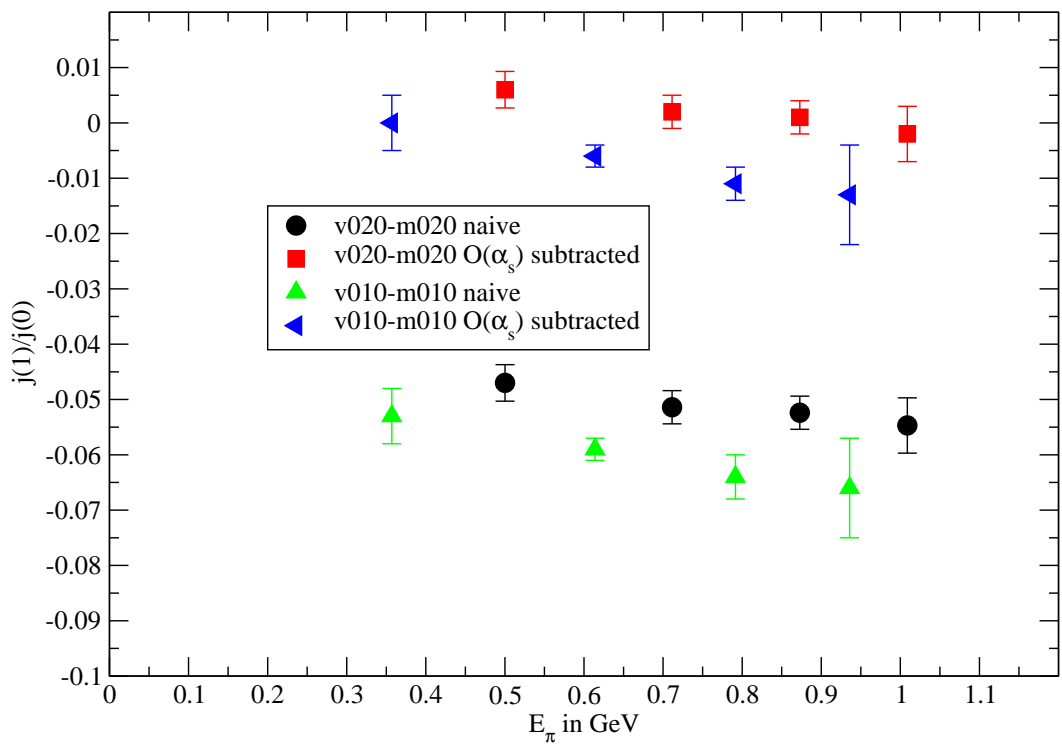}
\caption{The size of $1/M$ current correction $J_0^{(1)}$ is shown. The more physical subtracted currents have an even smaller effect than the naive ones, with the central value of the largest $J_0^{(1)}/J_0^{(0)}$ around 1 percent. }
\end{figure}

\section{Form Factors}
The semileptonic form factors are defined by the following relation with the hadronic matrix element of electroweak currents between a B meson and a pion:

\begin{eqnarray}
\langle \pi(p_\pi)| {V^\mu}| B(p_B) \rangle &=& 
{f_+(q^2) }
\, \left[ p_B^\mu + p_\pi^\mu - \frac{M_B^2-m_\pi^2}{q^2} 
\, q^\mu \right] \nonumber + {f_0(q^2)}
 \, \frac{M_B^2 - m_\pi^2}{q^2} \, q^\mu \nonumber \\
 &=& \sqrt{2 M_B} \,[v^\mu {f_\parallel}\,
 + \, p^\mu_\perp {f_\perp} ],  \nonumber
\end{eqnarray}
where {$v^\mu = \frac{p_B^\mu}{M_B}$,  $\; p_\perp^\mu 
=p_\pi^\mu - (p_\pi \cdot v) \, v^\mu$, $\; q^\mu = p_B^\mu - p_\pi^\mu$}.
\vspace{.6in}
\begin{figure}[h]
\includegraphics[width=14.cm]{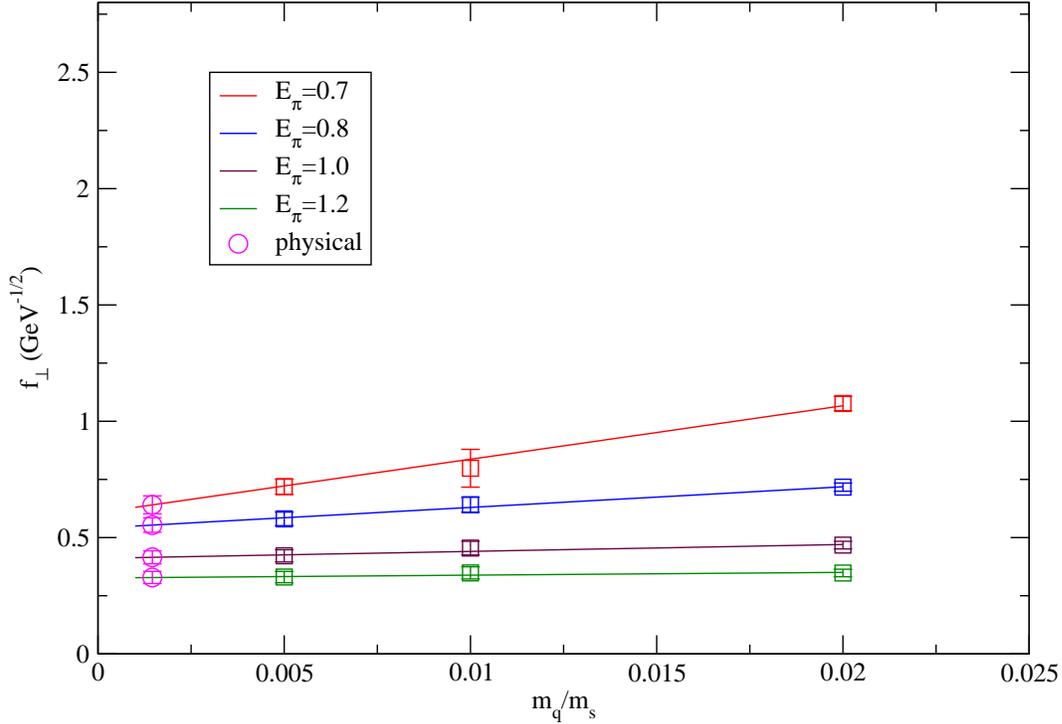}
\caption{Linear extrapolations of $f_{\perp}$ to the physical chiral limit at several fixed values of $E_\pi$.}
\end{figure}

The three-point correlator 
\begin{eqnarray}
C^{(3)}(\vec{p}_\pi, \vec{p}_B, t, T_B)   = \sum_{\vec{z}} \sum_{\vec{y}} { \langle }\Phi_\pi(0) J^\mu(\vec{z},t) \Phi^\dagger_B(\vec{y},T_B) { \rangle } e^{i\vec{p}_B\cdot \vec{y}} \, e^{i(\vec{p}_\pi - \vec{p}_B)\cdot \vec{z}}, \nonumber
\end{eqnarray}
\noindent
where $\Phi$'s are interpolating operators for the mesons, is fit using Bayesian fitting algorithms to the following form to obtain $A_{00}$:
\noindent
\begin{eqnarray}
C^{(3)}(\vec{p}_\pi, \vec{p}_B, t, T_B)  { \rightarrow }{ \sum_{k=0}^{N_\pi-1} \sum_{j=0}^{N_B-1}} (-1)^{k*t} (-1)^{j*(T_B-t)} { A_{jk}} { e^{-E_\pi^{(k)} t} e^{ -E_B^{(j)} (T_B-t)}}.
 \nonumber
\end{eqnarray}
$A_{00}$ is directly related to form factors $f_{\|}$ and $f_{\perp}$ in the following way:
\begin{eqnarray}
f_{\|} & = & \frac{{ A_{00}(V^0)}}{\sqrt{\zeta^{(0)}_\pi \zeta^{(0)}_B}}\sqrt{2 E_\pi},  \nonumber \\
f_{\perp} & = & \frac{{A_{00}(V^k)}}{\sqrt{\zeta^{(0)}_\pi \zeta^{(0)}_B} p_\pi^k }\sqrt{2 E_\pi},  \nonumber
\end{eqnarray}
where $A_{00}$ includes the effects of matching, and $\zeta^{(0)}_{B,\pi}$ are the groundstate amplitudes from the B meson and the pion $2-pnt$ correlators respectively.

\section{Procedure for Approaching the Chiral Limit}
\vspace{.6in}
\begin{figure}[h]
\includegraphics[width=14.cm]{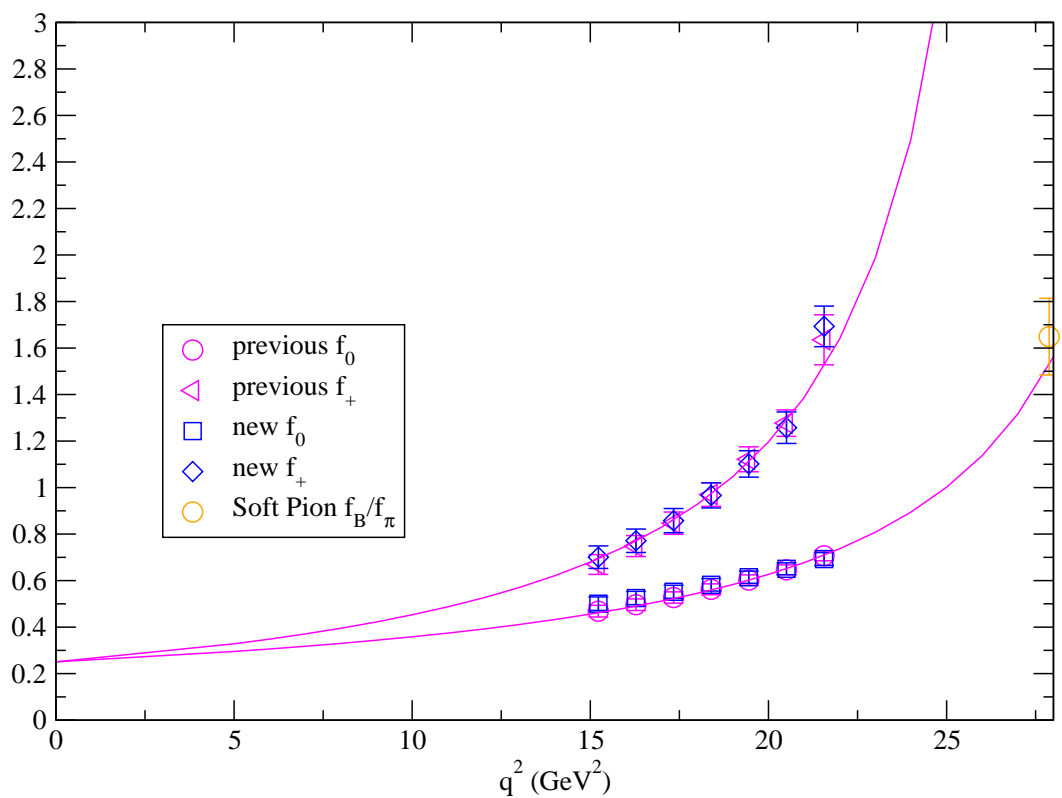}
\caption{ The form factors $f_+(q^2)$ and $f_0(q^2)$, calculated using the same ensemble as the previous year, including the $1/M$ current corrections are shown and compared with the results from last year. The curve shown is a BK \cite{BK} fit to last year's data. Results from other ensembles and more sophisticated chiral extrapolations have not been included yet in this plot. }
\end{figure}
One needs to extrapolate the form factors obtained as described in the previous section to the physical chiral limit. We first interpolate $f_{\|}$ and $f_{\perp}$ to the same values of $E_\pi$. After interpolating, we do chiral extrapolation at several fixed $E_\pi$ values to get the form factors at the physical point. A plot of $f_{\perp}$ showing the interpolated points at fixed $E_\pi$, and linear extrapolation to the physical limit can be seen in fig.2. So far, only simple linear fits were used in our results. We have started looking into doing more complicated chiral fits using \cite{xpt}. 


Using the chirally extrapolated $f_{\|}$ and $f_{\perp}$, we obtain $f_+(q^2)$ and $f_0(q^2)$. The form factors $f_+(q^2)$ and $f_0(q^2)$, calculated using the same ensemble as the previous year, including the $1/M$ current corrections are shown in Fig.3 and compared with the results from last year. The curve shown is a BK \cite{BK} fit to last year's data. Our results are in the region of $q^2$ larger than $15GeV$. We perform fits of our results to model ansatze for the $q^2$ dependence in order to obtain the form factors at smaller $q^2$. A new approach "Moving NRQCD" \cite{moving} shows promise to directly calculate the form factors even at small $q^2$. 
\section{Summary}
Last year, unquenched studies of semileptonic B decays using NRQCD heavy and improved staggered (AsqTad) light quark actions were reported. The main change since then is that we have included all 1/M corrections to currents through $O(\alpha_s,a \alpha_s, \alpha_s/aM,\alpha_s\Lambda_{QCD}/M)$, instead of only $O(\alpha_s)$. We also have more ensembles, instead of just one ensemble with dynamical quark mass $am=0.01$. We have started coding and examining the effects of staggered chiral perturbation theory. The results shown in this article, however, used only linear fits. We also have new data on finer lattice spacing that needs to be fit and included. Once our analysis is complete, and we have the corresponding $f_+(q^2)$ and $f_0(q^2)$, this will enable us to extract the CKM matrix element $|V_{ub}|$ using the new experimental  $B->\pi l \nu$ branching fraction results, for instance due to Belle, BaBar, CLEO.


\begin{thebibliography}{99}
  
  \bibitem{lastyear}
  J.\ Shigemitsu {\em et al.}; Nucl.\ Phys. B \ Proc.\ Suppl.\ {\bf 140},464 (2005), hep-lat/0408019.
  
  \bibitem{advisortalk}
  J.\ Shigemitsu; these proceedings.
  
  \bibitem{xpt}
  C.\ Aubin and C.\ Bernard; Staggered Chiral Perturbation Theory Notes; private communication.
  
  \bibitem{me}
  E.\ Gulez {\em et al.};
   Phys.\ Rev.\ {\bf D69},074501 (2004).

  \bibitem{BK} 
  D.\ Becirevic and A.\ B.\ Kaidalov;
   Phys.\ Lett.\ {\bf B478},417 (2000),  
  
  \bibitem{moving}
  K.\ Foley and G.\ P.\ Lepage;
   Nucl.\ Phys.\ Proc.\ Suppl.\ {\bf 119},635 (2003),\\
  K.\ Foley {\em et al.};Nucl.\ Phys. B \ Proc.\ Suppl.\ {\bf 140},470 (2005) \\
  A.\ Dougall; these proceedings

\end{thebibliography}
\end{document}